\documentstyle[aps,pra,eqsecnum,psfig,multicol,graphicx]{revtex}

\begin{document}
\draft
\title{Feedback-stabilization of an arbitrary pure state of a two-level atom}
\author{Jin Wang${}^{1}$, H.M. Wiseman${}^{2}$}
\address{${}^{1}$Centre for Laser Science,
Department of Physics, The University of Queensland, Brisbane,
Queensland 4072, Australia\\
${}^{2}$ School of Science, Griffith University, Brisbane, Queensland
4111, Australia \\}
\date{\today}
\maketitle

\begin{abstract}
Unit-efficiency homodyne detection of the resonance fluorescence of a
two-level atom collapses the quantum state of the atom to a stochastically
moving point on the Bloch sphere. Recently,
Hofmann, Mahler, and Hess [Phys.
Rev. A {\bf 57}, 4877 (1998)] showed
that by making part of the coherent driving
proportional to the homodyne photocurrent can stabilize the state to
any point on the bottom half of the sphere. Here we reanalyze their
proposal using the technique of stochastic master equations, allowing
their results to be generalized in two ways.
First, we show that any point on the upper or
lower half, but not the equator, of the sphere may be stabilized.
Second, we consider non-unit-efficiency detection, and quantify the
effectiveness of the feedback by calculating the maximal purity
obtainable in any particular direction in Bloch space.
\end{abstract}


\newcommand{\beq}{\begin{equation}}
\newcommand{\eeq}{\end{equation}}
\newcommand{\bqa}{\begin{eqnarray}}
\newcommand{\eqa}{\end{eqnarray}}
\newcommand{\nn}{\nonumber}
\newcommand{\nl}[1]{\nn \\ && {#1}\,}
\newcommand{\erf}[1]{Eq.~(\ref{#1})}
\newcommand{\rf}[1]{(\ref{#1})}
\newcommand{\dg}{^\dagger}
\newcommand{\rt}[1]{\sqrt{#1}\,}
\newcommand{\smallfrac}[2]{\mbox{$\frac{#1}{#2}$}}
\newcommand{\half}{\smallfrac{1}{2}}
\newcommand{\bra}[1]{\langle{#1}|}
\newcommand{\ket}[1]{|{#1}\rangle}
\newcommand{\ip}[2]{\langle{#1}|{#2}\rangle}
\newcommand{\sch}{Schr\"odinger }
\newcommand{\schs}{Schr\"odinger's }
\newcommand{\hei}{Heisenberg }
\newcommand{\heis}{Heisenberg's }
\newcommand{\bl}{{\bigl(}}
\newcommand{\br}{{\bigr)}}
\newcommand{\ito}{It\^o }
\newcommand{\str}{Stratonovich }
\newcommand{\dbd}[1]{\frac{\partial}{\partial {#1}}}
\newcommand{\sq}[1]{\left[ {#1} \right]}
\newcommand{\cu}[1]{\left\{ {#1} \right\}}
\newcommand{\ro}[1]{\left( {#1} \right)}
\newcommand{\an}[1]{\left\langle{#1}\right\rangle}
\newcommand{\implies}{\Longrightarrow}

\begin{multicols}{2}

\section{Introduction}
Although classical models of feedback schemes have been used for a
long time to
control dynamical noise, an analogous quantum theory of feedback has
been developed only in the last fifteen years
\cite{HauYam86,YamImoMac86,Sha87,WisMil93b,Wis94a,Pli94,DohJac99}.
Recently there has been considerable interest in quantum feedback
as a way to fight decoherence in isolated quantum systems, using the
approach of Refs.~\cite{WisMil93b,Wis94a}.
The central idea is to use a continuous measurement record, whose
existence is due to the coupling of the system to a bath, to control
the dynamics of the system so as to counteract the noise introduced by
that bath and possibly other baths.
For example, it has been suggested as
a way to create optical squeezed states \cite{WisMil94a}, to create
micromaser number states \cite{LieMil95}, to
correct errors in quantum bits \cite{MabZol96}, and to protect
optical and microwave \sch cat states against dissipation
\cite{TomVit94,HorKil97,VitTomMil97}.

Decoherence in quantum systems can be loosely defined as loss of
purity. Therefore the ultimate success in using feedback to
fight decoherence would
be to create an arbitrary stable pure state in the presence of dissipation.
This goal was realized (better even that they realized)
by Hoffman, Mahler and Hess (HMH) \cite{HofHesMah98,HofMahHes98} for
a very simple system: a resonantly driven two-level atom. They showed
that by using the photocurrent derived from unit-efficiency homodyne
detection of the atom's fluorescence to control part of the driving
field of the atom, it is possible to exactly cancel the noise
introduced by the electromagnetic vacuum field when the atom is in a
particular pure state. By choosing the
driving strength and feedback strength appropriately, any pure state
on the Bloch sphere may be picked out, although HMH claimed that only
pure states on the lower half of the sphere would be stable under
their scheme.

HMH chose to describe detection and feedback in their system in a way
different from (but equivalent to) the standard approach in
Refs.~\cite{WisMil93b,Wis94a}. In this paper we reformulate their
theory using the latter approach. This has the advantage of enabling
a number of generalizations of their results. First, we revisit the
question of stability and find that, contrary to the claims of HMH,
the states in the upper half of the Bloch sphere can be stabilized as
well as those in the lower half (this is what was better than they
realized). The only states which cannot be stabilized, in the sense
that an arbitrary initial state would not always end up in the desired
state, are those on the equator of the Bloch sphere; that is, those
which are equal superpositions of excited and ground states.

Our second generalization is to consider how effective feedback is with
$\eta < 1$; that is, with
non-unit-efficiency detection. In this case it is not possible to
stabilize the atom at any pure state, except the ground state which
is trivially stable by setting the driving and feedback to zero.
Instead, we aim to produce a steady state which is as close as
possible to a given pure state. For the two-level atom, this is
equivalent to trying to create a state which is as pure as possible in
a particular direction in Bloch space. Not surprisingly (given the
above result), we find that
states near the equator cannot be well-protected against decoherence.
We also find an echo of the distinction HMH found between the upper
and lower halves of the Bloch sphere, in that states in the upper half
sphere are affected much more by loss of detection efficiency that
those in the lower half.

The paper is organized as follows. In Sec.~II we present the model of
a driven two-level atom, including the stochastic \sch equation for
unit-efficiency homodyne detection. In Sec.~III we use this equation
to derive the driving and feedback required to stabilize the atom in
an arbitrary pure state. These results agree with those of HMH.
However, our stability analysis disagrees substantially
with theirs. In Sec.~IV we
present entirely new analytical results relating to the effect of non-unit-
efficiency detection. In Sec.~V we give numerical simulations of the
stochastic evolution equations, illustrating the issues discussed in
the preceding two sections.  In Sec.~VI we summarize and interpret our
results, explain their significance, and discuss the possibility of future
work.

\section{Homodyne Detection}

\subsection{Master Equation}

\begin{figure}[tbp]
\includegraphics[width=0.45\textwidth]{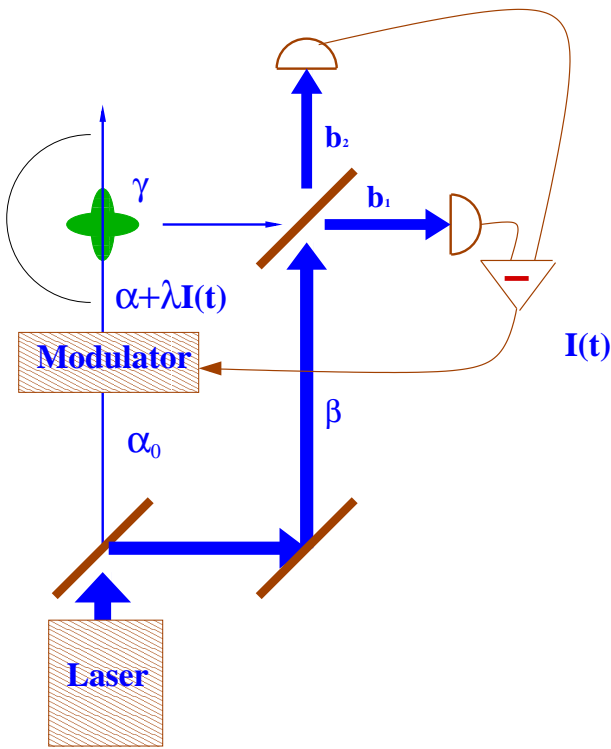}
\caption{\narrowtext Diagram of the experimental apparatus. The laser
beam is split to produce both the local oscillator $\beta$
and the field $\alpha_{0}$ which is modulated using the current $I(t)$.
The modulated beam,
with amplitude proportional to $\alpha+\lambda I(t)$, drives an atom
at the centre of the parabolic mirror. The fluorescence thus
collected is subject to homodyne detection using the local
oscillator, and gives rise to the homodyne photocurrent $I(t)$.}
	\protect\label{fig:diag}
\end{figure}

Consider a atom, with two relevant levels
$\{\ket{g},\ket{e}\}$ and lowering operator
$\sigma=\ket{g}\bra{e}$. Let the  the decay rate be $\gamma$, and let
it be driven by a resonant classical driving field with Rabi
frequency $2\alpha$. This is as shown in Fig.~\ref{fig:diag},
where for the moment
we are omitting feedback by setting $\lambda=0$.
This system is well-approximated by the master
equation
\beq \label{me1}
\dot{\rho} = \gamma{\cal D}[\sigma]\rho - i\alpha [\sigma_{y},\rho],
\eeq
where the Lindblad \cite{Lin76} superoperator is defined as usual
${\cal D}[A]B \equiv ABA\dg - \{A\dg A,B\}/2$.
In this master equation
 we have chosen to define the $\sigma_{x}=\sigma+\sigma\dg$ and
$\sigma_{y}=i\sigma-i\sigma\dg$ quadratures of the atomic dipole
relative to the driving field. The effect of driving is to rotate the
atom in Bloch space around the $y$-axis. The state of the atom in
Bloch space is described by the three-vector $(x,y,z)$. It is related
to the state matrix $\rho$ by
\beq \label{Bloch}
\rho = \frac{1}{2}\ro{I + x\sigma_{x}+y \sigma_{y}+z\sigma_{z}}.
\eeq

It is easy to show that the stationary solution of the master
equation (\ref{me1}) is
\bqa
{x}_{\rm ss}&=&\frac{4\alpha\gamma}{\gamma^2+8\alpha^2},\\
{y}_{\rm ss}&=&0         ,\\
{z}_{\rm ss}&=&\frac{-\gamma^2}{\gamma^2+8\alpha^2}.
\eqa
For $\gamma$ fixed, this is a family of solutions parameterized by the
driving strength $\alpha \in (-\infty,\infty)$. All members of the
family are in the $x$--$z$ plane on the Bloch sphere. Thus for this
purpose we can reparametrize the relevant states using $r$ and
$\theta$ by
\bqa
x &=& r\sin \theta \label{rtx}\\
z &=& r\cos \theta, \label{rtz}
\eqa
where $\theta \in [-\pi,\pi]$. Since
\beq
{\rm Tr}[\rho^{2}] = \frac{1}{2}\ro{1 + x^{2}+y^{2}+z^{2}}
\eeq
is a measure of the purity of the Bloch sphere, $r=\sqrt{x^{2}+z^{2}}$,
the distance from
the centre of the sphere, is also a measure
of purity. Pure states correspond to $r=1$ and maximally mixed
states to $r=0$.

\begin{figure}[tbp]
\includegraphics[width=0.45\textwidth]{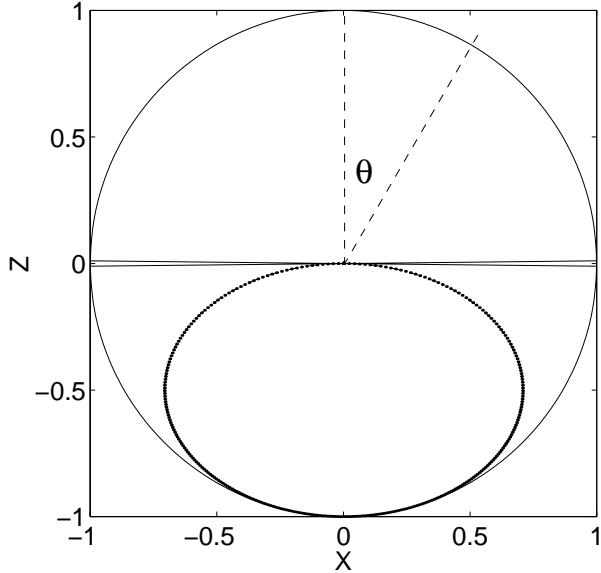}
\caption{\narrowtext Locus of the solutions to the Bloch equations.
The ellipse in the lower half plane is the locus for the equations
with driving only. The full circle (minus the points on the equator)
is the locus for the equations with optimal driving and feedback, as
defined in Sec.~III.}
	\protect\label{fig:bs1}
\end{figure}

The locus of
solutions in this plane (an ellipse) is shown in Fig.~\ref{fig:bs1}. Since
$z_{\rm ss}<0$, all solutions are in the lower half of the Bloch sphere.
That is, we are restricted to $|\theta| > \pi/2$. Also, it is evident
that the smaller $|\theta|$ is (that is, the more excited the atom is),
the smaller $r$ is (that is, the less pure the atom is). At
$|\theta|=\pi$ the stationary state is pure, but this is not
surprising as it is simply the ground state of the atom with no
driving. As
$|\theta|\to \pi/2$ we have $r \to 0$. This can only be approached
asymptotically as $|\alpha| \to \infty$.
In summary, the stationary states we can reach by driving
the atom are limited, and generally far from pure.

\subsection{Homodyne Detection}

Now consider subjecting the atom to homodyne detection. As shown in
Fig.~\ref{fig:diag}, we assume that all of the fluorescence of the
atom is collected and turned into a beam (represented in
Fig.~\ref{fig:diag} by
placing the atom at the focus of a mirror). Ignoring the vacuum
fluctuations in the field, the annihilation operator for this beam is
$\sqrt{\gamma}\sigma$, normalized so that the mean intensity
$\gamma\an{\sigma\dg \sigma}$ is equal to
the number of photons per unit time in the beam.
This beam then enters one
port of a 50:50 beam splitter, while a strong local oscillator
$\beta$ enters the other. To ensure that this local oscillator has a
fixed phase relationship with the driving laser used in the measurement, it
would be
natural to utilize the same coherent light field source in the
driving process
and as the local oscillator in the homodyne detection. This is as
shown in Fig.~\ref{fig:diag}.

Again ignoring vacuum fluctuations, the two field operators exiting
the beam splitter, $b_{1}$ and $b_{2}$, are
\beq
b_{k} = \sq{\sqrt{\gamma}\sigma -(-1)^{k}\beta}/\sqrt{2}.
\eeq
When these two fields are detected, the two photocurrents produced
have means
\beq
\bar{I}_{k} = \an{ |\beta|^{2} - (-1)^{k}\ro{ \sqrt{\gamma}\beta\sigma\dg +
\sqrt{\gamma}\sigma \beta^{*}} + \gamma\sigma\dg\sigma}/2.
\eeq
The middle two terms represent the interference
between the system and the local
oscillator.

The ideal limit of homodyne detection is when the local
oscillator amplitude
goes to infinity, which in practical terms means
$|\beta|^2\gg \gamma $. In this limit,
the rate of the photodetections goes
to infinity and thus it should be possible to change the point process of
photocounts into a continuous photocurrent with white noise. Also,
the only relevant quantity is the difference between the two photocurrents.
Suitably normalized, this is \cite{Car93b,WisMil93a}
\beq \label{homo1}
I(t) = \frac{I_{1}(t)-I_{2}(t)}{|\beta|} =
\sqrt{\gamma}\an{e^{-i\Phi}\sigma\dg + e^{i\Phi}\sigma}_{\rm c}(t) + \xi(t).
\eeq

A number of aspects of \erf{homo1} need to be explained. First,
$\Phi = \arg\beta$, the phase of the local oscillator (defined relative to the
driving field). Second, the subscript c means conditioned and refers
to the fact that if one is making a homodyne measurement then this
yields information about the system. Hence, any system averages will
be conditioned on the previous photocurrent record. Third, the final
term $\xi(t)$ represents Gaussian white noise, so that
\beq
\xi(t)dt = dW(t),
\eeq
an infinitesimal Wiener increment defined by \cite{Gar85}
\bqa
dW(t)^2=dt , \label{ito1}\\
{\rm E}[dW(t)]=0 . \label{dW0}
\eqa

Since the stationary solution of the master equation
confines the state to the $x$--$z$ plane, it makes sense to follow HMH
by setting $\Phi=0$. In that case,
\beq \label{homo2}
I(t) =\sqrt{\gamma}\an{\sigma_{x}}_{\rm c}(t) + \xi(t).
\eeq
That is,
the deterministic part of the
homodyne photocurrent is proportional to $x_{\rm c} =
\an{\sigma_{x}}_{\rm c}$. This should be useful for controlling
the dynamics of the state in the $x$--$z$ plane by feedback, as we
will consider in Sec.~III. Of course, all that really matters here is
the relationship between the driving phase and the local oscillator
phase, not the absolute phase of either.

The conditioning process referred to above can be made explicit by
calculating how the system state changes in response to the measured
photocurrent. Assuming that the state at some point in time is pure
(which will tend to happen because of the conditioning anyway), its
future evolution can be described by the stochastic \sch equation (SSE)
\cite{Car93b,WisMil93a}
\beq \label{SSE1}
d\ket{\psi_{\rm c}(t)} = \hat{A}_{\rm c}(t)\ket{\psi_{\rm c}(t)}dt
+ \hat{B}_{\rm c}(t)\ket{\psi_{\rm c}(t)}dW(t).
\eeq
This is an \ito stochastic equation \cite{Gar85}
with a drift term and a diffusion
term. The operator for the drift term is
\beq
\hat{A}_{\rm c}(t) = \frac{\gamma}{2}\sq{-\sigma\dg\sigma
+ \an{\sigma_{x}}_{\rm c}(t)\sigma - \an{\sigma_{x}}_{\rm c}^{2}(t)/4}
-i\alpha\sigma_{y},
\eeq
while that for the diffusion is
\beq
\hat{B}_{\rm c}(t) = \sqrt{\gamma}\sq{\sigma - \an{\sigma_{x}}_{\rm
c}(t)/2}.
\eeq
Both of these operators are conditioned in that they depend on the
system average
\beq
\an{\sigma_{x}}_{\rm c}(t) = \bra{\psi_{\rm c}(t)}\sigma_{x}
\ket{\psi_{\rm c}(t)}.
\eeq

On average, the system still obeys the master equation (\ref{me1}).
This is easiest to see from the stochastic master equation
(SME), which allows for impure initial conditions. The SME can be
derived from the SSE by constructing
\bqa
d \ro{\ket{\psi_{\rm c}}\bra{\psi_{\rm c}}} &=&
\ro{d\ket{\psi_{\rm c}}}\bra{\psi_{\rm c}} +
\ket{\psi_{\rm c}}\ro{d\bra{\psi_{\rm c}}} \nl{+}
\ro{d\ket{\psi_{\rm c}}}\ro{d\bra{\psi_{\rm c}}},
\eqa
using the \ito rule (\ref{ito1}),
and then identifying $\ket{\psi_{\rm c}}\bra{\psi_{\rm c}}$ with
$\rho_{\rm c}$. The result is
\beq \label{SME1}
d\rho_{\rm c} = dt\gamma{\cal D}[\sigma]\rho_{\rm c} - idt\alpha
[\sigma_{y},\rho_{\rm c}] +
dW(t)\sqrt{\gamma}{\cal H}[\sigma]\rho_{\rm c} ,
\eeq
where ${\cal H}[A]B \equiv AB+BA\dg - {\rm Tr}[AB+BA\dg]$. Although
this has been derived assuming pure initial conditions, it is valid
for any initital conditions \cite{WisMil93a}. This is
also an \ito equation, which means the evolution for the
ensemble average state matrix
\beq
\rho(t)={\rm E}[\rho_{\rm c}(t)]
\eeq
is found simply by averaging over the photocurrent
noise term by using \erf{dW0}.
This procedure yields the original master equation (\ref{me1}) again.
The general term for the stochastic conditioned evolution of the
system, be it described by a SSE or SME, is a quantum trajectory
\cite{Car93b}, and the quantum trajectory is said to unravel the
master equation \cite{Car93b}.

\section{Feedback with unit-efficiency detection}

\subsection{SSE including feedback} \label{sec:SSE}

We now include feedback onto the amplitude of the driving on the atom,
proportional to the homodyne photocurrent, as done by HMH.
This is as shown in Fig.~\ref{fig:diag}, where the driving field
passes through an electro-optic
amplitude modulator controlled by the photocurrent, yielding a field
proportional to $\alpha + \lambda I(t)$. This means that the feedback
can be described by the Hamiltonian
\beq \label{fbH}
H_{\rm fb} = \lambda \sigma_{y}I(t).
\eeq
In this paper we are assuming instantaneous feedback, where the time
delay in the feedback loop is negligible.

Since the homodyne photocurrent (\ref{homo1}) is defined in terms of
system averages and the noise $dW(t)$, the SSE including feedback can
still be written as an equation of the form (\ref{SSE1}). The
effect of the
feedback Hamiltonian can be shown \cite{WisMil93b,WisMil94a} to
change the drift and diffusion operators to
\bqa
\hat{A}_{\rm c}(t) &=& \frac{\gamma}{2}\sq{-\sigma\dg\sigma
+ \an{\sigma_{x}}_{\rm c}(t)\sigma - \an{\sigma_{x}}_{\rm c}^{2}(t)/4}
-i\alpha\sigma_{y}
\nl{+} \lambda \sqrt{\gamma}
\sq{-i \an{\sigma_{x}}_{\rm c}(t)\sigma_{y} -2\sigma\dg \sigma}  -
{\lambda^{2}}/{2}, \\
\hat{B}_{\rm c}(t) &=& \sqrt{\gamma}\sq{\sigma - \an{\sigma_{x}}_{\rm
c}(t)/2} - i\lambda\sigma_{y}.
\eqa

Say we wish to stabilize the pure state with Bloch angle $\theta$,
as defined in Eqs.~(\ref{rtx}) and (\ref{rtz}), with $r=1$ of course.
In terms of the ground and excited states, this state is
\beq
 \ket{\theta}=\cos\frac{\theta}{2}\ket{e}+\sin\frac{\theta}{2}\ket{g}.
\eeq
Now for this state to be stabilized we must have
\beq
\sq{\hat{A}_{\rm c}(t)dt + \hat{B}_{\rm c}(t)dW(t)}\ket{\theta}
\propto \ket{\theta}.
\eeq
We cannot say the left-hand-side should equal zero because a change in
the overall phase still leaves the physical state unchanged. However,
we can work with this equation, and simplify it by dropping all terms
proportional to the identity operator in $\hat{A}_{\rm c}(t)$ and
$\hat{B}_{\rm c}(t)$. We can also demand that it be satisfied for the
deterministic and noise terms separately, because $dW(t)$ can take any
value. This gives the two equations
\bqa
\ro{ \sqrt{\gamma}\sigma -i\lambda\sigma_{y}}\ket{\theta} &\propto&
\ket{\theta} ,\\
\left[\gamma\ro{-\sigma\dg\sigma
+ \sin\theta \sigma}
-i2\alpha\sigma_{y}\phantom{\ket{\sqrt{\gamma}}}\right. && \nn \\
\left. +\, \lambda \sqrt{\gamma}
\ro{-i \sin\theta\sigma_{y}/2 -\sigma\dg \sigma}\right] \ket{\theta}
&\propto& \ket{\theta},
\eqa
where we have put $\an{\sigma_{x}}_{\rm c}(t)$ equal to $\sin\theta$,
its value for the state $\ket{\theta}$.

\begin{figure}[tbp]
\includegraphics[width=0.45\textwidth]{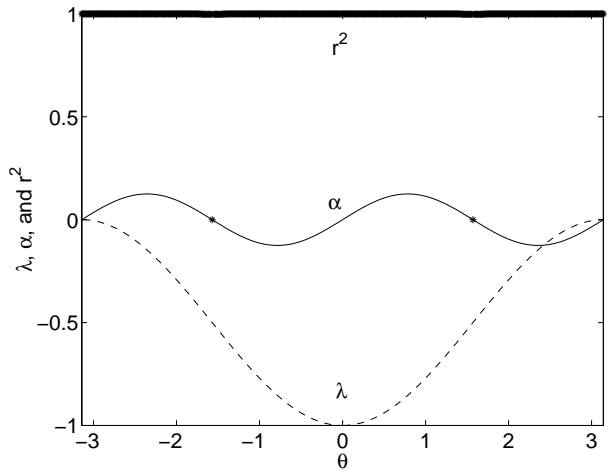}
\caption{\narrowtext Plot of the optimal driving ($\alpha$, solid)
and feedback ($\lambda$, dashed) required to produce a pure state
with Bloch angle $\theta$. For this plot we have set $\gamma=1$ so
that $\alpha$ and $\lambda$ are dimensionless.
The purity ($r^{2}$, starred) is one except
for $\theta=\pm \pi/2$, where the feedback is not stable.}
	\protect\label{fig:eta1fns}
\end{figure}

Solving the first equation easily yields the condition
\beq \label{lth}
\lambda = -\frac{\sqrt{\gamma}}{2}(1+\cos\theta).
\eeq
This is equivalent to the feedback condition derived by HMH, stated
as Eq.~(35) of Ref.~\cite{HofMahHes98}.
Substituting this into the second equation gives, after some
trigonometric manipulation, the second condition
\beq \label{ath}
\alpha = \frac{\gamma}{4}\sin\theta\cos\theta.
\eeq
Again, this agrees with the driving strength of HMH, given as Eq.~(44)
of Ref.~\cite{HofMahHes98}. It is worth emphasizing that the derivation
given here is entirely different in detail from that of HMH, and so
is an
independent verification of their result. These functions are plotted
in Fig.~\ref{fig:eta1fns}. Note that there are two points with the
same values of both $\lambda$ and $\alpha$, at $\theta = \pm \pi/2$.

\subsection{Stability}

The preceding derivation seems to show that any pure state can be
stabilized by a suitable choice of driving and feedback.
Indeed our derivation proves that that if one prepares a
state in exactly the pure state one desires, then the feedback scheme
of HMH which we have analyzed will keep the system in that state.
However, to discuss stability we need to know what will happen for
states which are not initially in the desired state. To deal with this
it is much more convenient to use the SME rather than the SSE, as will
become apparent.

The SME can be constructed from the SSE in the same way as before. The
result is \cite{WisMil93b,WisMil94a}
\bqa
d\rho_{\rm c} &=&  dt\gamma{\cal D}[\sigma]\rho_{\rm c} - idt\alpha
[\sigma_{y},\rho_{\rm c}] \nl{-}
idt \lambda [\sigma_{y},\sigma\rho_{\rm c} + \rho_{\rm c}\sigma\dg] +
dt(\lambda^{2}/\gamma){\cal D}[\sigma_{y}]\rho_{\rm c} \nl{+}
dW(t){\cal H}[\sqrt{\gamma}\sigma-i\lambda\sigma_{y}]\rho_{\rm c}.
\eqa
Also as before, this is an \ito stochastic equation, which means that
the ensemble average can be found simply by dropping the stochastic
terms. This time, the result is not the original master equation, but
rather the feedback-modified master equation
\beq
\dot{\rho} = -i[\alpha\sigma_{y},\rho]+{\cal
D}[\sqrt{\gamma}\sigma-i\lambda\sigma_{y}]\rho \equiv {\cal L}\rho.
\eeq
Here we have put the Liouvillian superoperator ${\cal L}$ in
 a manifestly Lindblad form.

Now we have shown already that the pure state $\rho =
\ket{\theta}\bra{\theta}$ must be a
solution of this master equation, for the appropriate choices of
$\lambda$ (\ref{lth}) and $\alpha$ (\ref{ath}). But for it to be
a stable solution we require all of the eigenvalues of the resulting
${\cal L}_{\theta}$ to
have a negative real part (except for the one eigenvalue that is
always zero, as required for ${\cal L}_{\theta}$ to be norm-preserving).
It is not difficult to find these eigenvalues, and in terms of $\theta$
they are
\beq \label{evalues}
-{\gamma}/{2},-{\gamma}/{2},-\gamma\cos^{2}\theta.
\eeq
Evidently the state $\ket{\theta}$ will be stable for all $\theta$
except $\theta = \pm \pi/2$. That is, all states are stable except
those on the equator. This is contrary to the conclusion of HMH
\cite{HofMahHes98},
based on a linearized stability analysis, that ``long term stability
of \ldots inverted states [i.e. states in the upper half plane]
cannot be achieved.'' We
emphasize that our stability analysis contains no approximations.

In hindsight, the lack of stability for pure states on the equator
could have been predicted from the expressions (\ref{lth}) and
(\ref{ath}). As discussed above and
shown in Fig.~\ref{fig:eta1fns}, the values for driving and feedback
for  $\theta=\pi/2$ are the same as those for $\theta=-\pi/2$. This
means that both $\rho=\ket{\pi/2}\bra{\pi/2}$ and
$\rho=\ket{-\pi/2}\bra{-\pi/2}$
are solutions of ${\cal L}_{\theta}\rho=0$ for $\theta=\pi/2$ {\em or}
$-\pi/2$. By linearity, any mixture of $\ket{\pi/2}\bra{\pi/2}$ and
$\ket{-\pi/2}\bra{-\pi/2}$ will be a solution also. Hence any
deviation away from one of these pure state will not necessarily be
suppressed, and the system lacks stability. With random
external perturbations, the
system will eventually reach an equal mixture of $\ket{\pi/2}\bra{\pi/2}$ and
$\ket{-\pi/2}\bra{-\pi/2}$, which is a state with $r=0$ (minimum
purity). This is
why we have plotted a value of $r=0$ in Fig.~\ref{fig:eta1fns} for
$|\theta|=\pi/2$.  We also plot $r$ as a function of $\theta$ in
Bloch space in Fig~\ref{fig:bs1}, giving the locus of states which
can be stabilized by feedback. This can compared to the locus of the
mixed states which
are accessible by driving alone. We will return to the stability
issue in the context of stochastic dynamics in Sec.~V.

\section{Feedback with non-unit-efficiency detection}

We have seen that the stochastic master equation is a very useful
representation of a quantum trajectory, as it allows the
unconditioned (deterministic) master equation to be derived easily,
and this latter equation is all that is required for a completely
rigorous stability analysis. The SME is also superior to the SSE in
that allows inefficient detection to be described. In a real
experiment this has to be taken into account. The effect of non-unit
$\eta$ on feedback in the present system is of interest both in
itself, and because of the extreme nonlinearity of the system dynamics
as compared to other quantum optical feedback systems such as
considered in Ref.~\cite{WisMil94a}.

As explained in
Ref.~\cite{WisMil93a}, the homodyne photocurrent
from a detection scheme with efficiency $\eta$ is
\beq \label{homo3}
I(t) = \sqrt{\gamma}\an{\sigma_{x}}_{\rm c}(t) +
\xi(t)/\sqrt{\eta}.
\eeq
Here we have used a normalization such that the deterministic part does
not depend on $\eta$. The effect of decreased efficiency is
increased noise. This means that we can retain the same feedback
Hamiltonian as above (\ref{fbH}), without changing the significance
of the feedback parameter $\lambda$. The SME with $\eta<1$, including
feedback, is \cite{WisMil94a}
\bqa
d\rho_{\rm c} &=&  dt\gamma{\cal D}[\sigma]\rho_{\rm c} - idt\alpha
[\sigma_{y},\rho_{\rm c}] \nl{-}
idt \lambda [\sigma_{y},\sigma\rho_{\rm c} + \rho_{\rm c}\sigma\dg] +
dt(\lambda^{2}/\gamma\eta){\cal D}[\sigma_{y}]\rho_{\rm c} \nl{+}
dW(t){\cal H}[\sqrt{\gamma\eta}\sigma-i\lambda\sigma_{y}/\sqrt{\eta}]
\rho_{\rm c} \label{SME2}
\eqa
The no-feedback SME, analogous to \erf{SME1}, can be obtained simply by
setting $\lambda = 0$, and was derived in Ref.~\cite{WisMil93a}.

Once again, it is easiest for the moment to just consider the ensemble
average evolution by averaging $dW$ to zero. The Lindblad form of the
resulting master equation is
\beq \label{me3}
\dot{\rho} = -i[\alpha\sigma_{y},\rho]+{\cal
D}[\sqrt{\gamma}\sigma-i\lambda\sigma_{y}]\rho +
(\lambda^{2}/\eta){\cal D}[\sigma_{y}]\rho.
\eeq
We do not know {\em a priori} what values of $\lambda$ and $\alpha$ to
choose to give the best results with inefficient detection, as the
SSE analysis in Sec.~\ref{sec:SSE} obviously does not apply. Hence we
simply solve for the stationary matrix in terms of $\alpha$ and
$\lambda$. Using the Bloch representation we find
\bqa
{x}_{\rm ss}&=&{4\alpha\eta^2(\gamma+2\sqrt{\gamma}\lambda)}/D,
\label{xss2}\\
{y}_{\rm ss}&=&0 ,\\
{z}_{\rm ss}&=&-\sqrt{\gamma}\eta(\sqrt{\gamma}+2\lambda)
(\gamma\eta+4\sqrt{\gamma}\eta\lambda+4\lambda^2)/D, \label{zss2}
\eqa
where
\bqa
D &=&
\gamma^2\eta^2+6\gamma^{3/2}\eta^{2}\lambda+2\gamma\eta(3+4\eta)\lambda^2+
16\sqrt{\gamma}\eta\lambda^3
\nl{+}8(\alpha^2\eta^2+\lambda^4).
\eqa

The question now arises, what do we mean by ``best results'' for the
feedback system. We cannot hope anymore to produce stable pure
states anywhere on the Bloch sphere. However, we can pick a
direction $\theta$ on the Bloch sphere and ask how close can we get to
a pure state? That is, we use the radius $r$ in \erf{rtx} and
\erf{rtz} as the quantity to be maximized, for each $\theta$. From
these two equations we have
\beq
\tan\theta = x_{\rm ss}/z_{\rm ss}.
\eeq
>From Eqs.~\rf{xss2} and \rf{zss2} we can immediately find the
desired driving in terms of $\lambda$ and $\theta$ as
\beq
\alpha = \frac{-\sqrt{\gamma}\eta(\sqrt{\gamma}+2\lambda)
(\gamma\eta+4\sqrt{\gamma}\eta\lambda+4\lambda^2)\tan\theta}
{4\eta^2(\gamma+2\sqrt{\gamma}\lambda)}. \label{alpha}
\eeq
The aim is then, for each $\theta$, to find the feedback $\lambda$ which
maximizes
\beq \label{minr}
r^{2} = x_{\rm ss}^{2}+z_{\rm ss}^{2}.
\eeq
This was done numerically using {\sc matlab}.

\begin{figure}[tbp]
\includegraphics[width=0.45\textwidth]{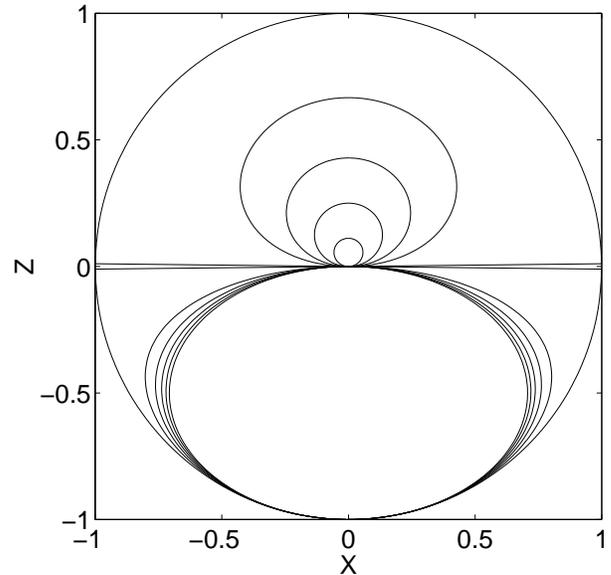}
\caption{\narrowtext Locus of the solutions to the Bloch equations
with optimal feedback for different values of detector efficiency
$\eta$. from the outside in, we have $\eta = 1, 0.8, 0.6, 0.4, 0.2, 0$. }
	\protect\label{fig:bs2}
\end{figure}

The results of our calculations are shown in Fig.~\ref{fig:bs2}, where
we plot the locus in Bloch space
of the best (most pure) stationary states which can
be achieved by feedback from non-unit-efficiency detection. We use a
variety of values of $\eta$.
A number of points are worth noting. First, and most obviously, the
degree of purity (measured by the $r$, the distance
from the origin)
decreases with $\eta$. Second, the gap at the equator for $\eta=1$
quickly widens, so that the purity of the best states with
$\theta$ close to $\pi/2$ is small. Third, the purity of the best
states in the upper half of the Bloch sphere is affected much more by
loss of detection efficiency than those in the lower half. Fourth, in
the limit $\eta=0$, the best solutions correspond to the no-feedback
solutions shown in Fig.~\ref{fig:bs1}. This is not surprising, since
with $\eta=0$ the photocurrent contains no information about the
system (as the noise is infinitely large) and hence there is no point
doing feedback. Since the stationary states with no feedback are
confined to the lower half of the Bloch sphere, this explains why
the best states with feedback in the lower half
 are less affected as $\eta$ decreases
than those in the upper half.

\begin{figure}[tbp]
\includegraphics[width=0.45\textwidth]{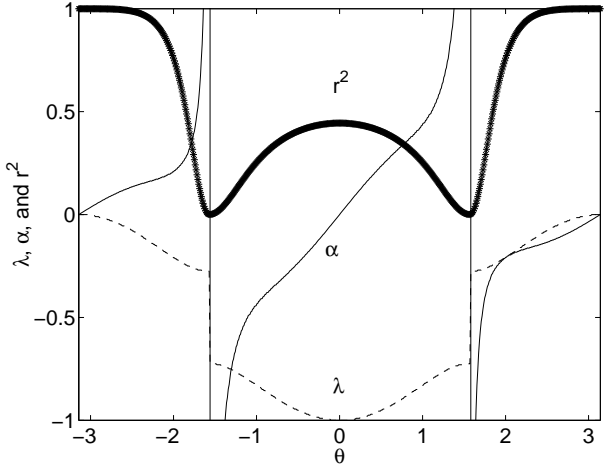}
\caption{\narrowtext Plot of the optimal driving ($\alpha$, solid)
and feedback ($\lambda$, dashed) required to produce the most pure state
with Bloch angle $\theta$. For this plot we have set $\gamma=1$ so
that $\alpha$ and $\lambda$ are dimensionless.
The purity obtain is also plotted ($r^{2}$, starred).}
	\protect\label{fig:etalt1fns}
\end{figure}

For the particular value $\eta=0.8$ we plot in
Fig.~\ref{fig:etalt1fns} the
values of $\alpha$ and $\lambda$ (as well as
purity, quantified as $r^{2}$) versus $\theta$. By comparing this plot with
Fig.~\ref{fig:eta1fns} one obtains some idea of the effect of
inefficient detection. A number of features remain the same.
First, $\alpha$ is antisymmetric in $\theta$, while $\lambda$ is symmetric.
Recalling that the deterministic part of the feedback is proportional
to $\lambda\an{\sigma_{x}}_{\rm c}$, the feedback itself is actually
antisymmetric as well as the driving. Second, the magnitude of the
feedback is zero for $|\theta|=\pi$ (the ground state) and increases
monotonically to a maximum of $\sqrt{\gamma}$ at $\theta=0$ (in the
direction of
the excited state). Third,
the driving is zero at the ground state  and at $\theta=0$, and
also changes sign as one passes through the equatorial place.
The most obvious difference between the parameters for $\eta=1$ and
those for
$\eta=0.8$ parameter is that the latter have a discontinuity at
$|\theta|=\pi/2$. The feedback parameter $\lambda$ jumps as one
crosses the
equatorial plane, while the driving $\alpha$ asymptotes to $+\infty$
on one side and $-\infty$ on the other. These extreme variations in
the driving do not prevent the best purity from approaching zero in the
equatorial plane.

\section{Stochastic Dynamics}

\subsection{Stochastic Bloch Equations}

So far we have considered the stochastic conditioned dynamics for the
system state in
order to derive the parameters $\lambda$ and $\alpha$ such that for
$\eta=1$ those dynamics are banished in the steady state. In this
section we will consider them in more detail, highlighting the
difference between the $\eta=1$ case and the $\eta <1$ case, and also
looking in more detail at the special case of $|\theta|=\pi/2$. The
most convenient way to treat the stochastic dynamics in general is
through the stochastic Bloch equations (SBE). These are simply the
stochastic equations for the conditioned Bloch vector, defined by
\beq
\rho_{\rm c} =  \frac{1}{2}\ro{I + x_{\rm c}\sigma_{x} +
y_{\rm c} \sigma_{y} + z_{\rm c}\sigma_{z}}.
\eeq

>From the SME including feedback (\ref{SME2}), we find
\bqa \left(\begin{array}{c}
dx_{\rm c}\\dy_{\rm c}\\dz_{\rm c}\end{array}\right)
&=& dt \left(\begin{array}{ccc}
-\gamma/2-2\kappa &0  &  2\alpha \\
0   & -{\gamma}/{2} & 0       \\
-2\alpha  & 0 &-\gamma-2\kappa\\
\end{array}\right)
\left(\begin{array}{c}
x_{\rm c}\\y_{\rm c}\\z_{\rm c}\end{array}\right)
\nl{-}dt\left(\begin{array}{c}
0\\0\\2\lambda\sqrt{\gamma}+\gamma\end{array}\right)
+ dW(t)\times \nn \\
&& \left(\begin{array}{c}
-\sqrt{\gamma\eta}x_{\rm c}^2+(\sqrt{\gamma\eta}+
{2\lambda}/{\sqrt{\eta}})z_{\rm c}+\sqrt{\gamma\eta}\\
-\sqrt{\gamma\eta}x_{\rm c}y_{\rm c}\\
-(\sqrt{\gamma\eta}+{2\lambda}/{\sqrt{\eta}})x_{\rm
c}-\sqrt{\gamma\eta}x_{\rm
c}z_{\rm c}\end{array}\right), \nn \\ \label{SBE1}
\eqa
where $\kappa = {\lambda^2}/{\eta}+\lambda\sqrt{\gamma}$. If we
ignore the final (noise) term, we get the Bloch equations from the
master equation (\ref{me3}).

\subsection{unit-Efficiency}

In the case $\eta=1$, considered in Sec.~III,
both the deterministic and stochastic dynamics disappeared
in the steady state for the appropriate choice of $\alpha$ and
$\lambda$.  Because the
stationary solution of the SSE was a unique pure state, that was
necessarily also the stationary solution of
the master equation found by averaging over the noise in the
equivalent SME. Thus
there was no distinction between the unconditioned and conditioned
states. There are two exceptions to this lack of distinction. The first is in
the transients, before the system reaches its steady state. The second
is for the special case $|\theta|=\pi/2$. In this subsection we
investigate these exceptions.

\begin{figure}[tbp]
\includegraphics[width=0.45\textwidth]{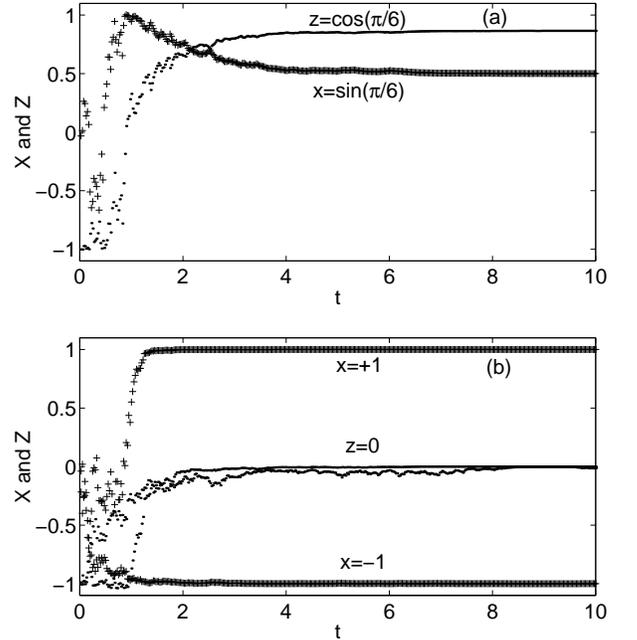}
\caption{\narrowtext Typical quantum trajectories for optimal feedback with
$\eta=1$, shown by $x_{\rm c}$ (plusses) and $z_{\rm c}$ (dots) as
functions of time. (a) shows a single trajectory for $\theta=\pi/6$, and (b)
two trajectories for $\theta=\pi/2$. }
	\protect\label{fig:traj}
\end{figure}

\begin{figure}[tbp]
\includegraphics[width=0.45\textwidth]{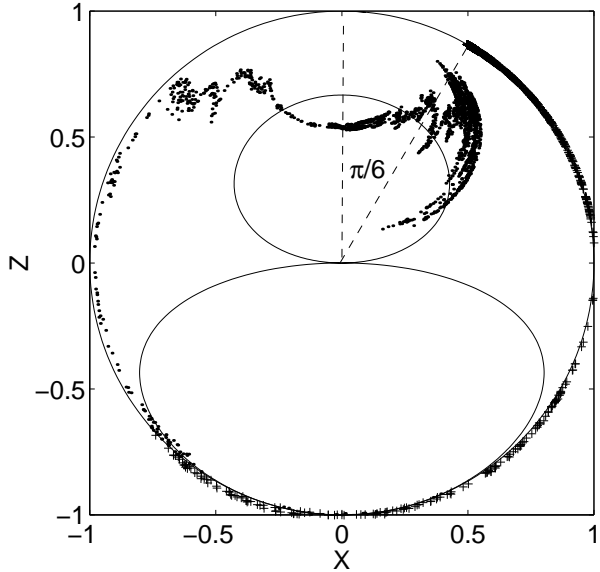}
\caption{\narrowtext Typical quantum trajectories in Bloch space for
$t\in [0,10\gamma^{-1}]$ under optimal feedback for
$\theta=\pi/6$, starting at the ground state.
The plusses are for $\eta=1$ and the dots for
$\eta=0.8$. The locus for the deterministic stationary states for
$\eta=0.8$ are also shown; the relevant state for this quantum
trajectory is at the intersection of the locus and the ray at
$\theta=\pi/6$. Note that the quantum trajectory for $\eta=0.8$
wanders around this average position,
while that for $\eta=1$ stops precisely at the
desired pure steady state.}
	\protect\label{fig:bs3}
\end{figure}

First, the transient behaviour was simulated using the SBE with
$\eta=1$. We chose the initial state to be the ground state, and
evolved the system stochastically from $t=0$ to $t=10\gamma^{-1}$.
With this choice of initial condition, $y_{\rm c}=0$ for all time.
We verified that in each trajectory $x_{\rm c}^{2}+z_{\rm c}^{2}=1$ to
a good approximation (indicating a pure state), but that the ensemble
averages over many trajectories
\beq
x = {\rm E}[x_{\rm c}]\;,\;\;z={\rm E}[z_{\rm c}]
\eeq
obey the deterministic Bloch equations.
A typical trajectory for $\theta=\pi/6$ is shown in
Fig.~\ref{fig:traj}(a). We see that the initial evolution is very erratic,
but that on a time scale of a few
$\gamma^{-1}$ the system relaxes towards a steady state which is pure
and stationary. By $t=10\gamma^{-1}$ the system is locked in a stable
pure state for all intents. We have also illustrated another typical
trajectory in Bloch space, in Fig.~\ref{fig:bs3}.

It is easy to verify that by putting $\eta=1$ and
\beq
(x_{\rm c},y_{\rm c},z_{\rm c}) = (\sin\theta,0,\cos\theta)
\eeq
in the right-hand-side of
the SBE (\ref{SBE1}), we obtain complete cancellation.
If we wish, we can follow HMH and
separate the noise term into the contribution from
feedback (proportional to $\lambda$) and that contribution present even
without
feedback (the rest). We interpret the latter stochasticity as being due
to the quantum measurement we are making, with its underlying
probabilistic nature. Obviously the fluctuation due to measurement is
canceled by the feedback, as HMH point out. It is equally important
that the deterministic dynamics are also canceled at this point.

The story for the special case $\theta=\pi/2$ is quite different.
 For this case the SBEs are
\bqa \left(\begin{array}{c}
dx_{\rm c}\\dy_{\rm c}\\dz_{\rm c}\end{array}\right)
&=& dt \left(\begin{array}{ccc}
0 &0  &  0 \\
0   & -{\gamma}/{2} & 0       \\
0  & 0 &-\gamma/2\\
\end{array}\right)
\left(\begin{array}{c}
x_{\rm c}\\y_{\rm c}\\z_{\rm c}\end{array}\right)
\nl{+}\sqrt{\gamma} dW(t)\left(\begin{array}{c}
1-x_{\rm c}^2\\
-x_{\rm c}y_{\rm c}\\
-x_{\rm c}z_{\rm c}\end{array}\right).
\eqa
Here the three eigenvalues in \erf{evalues} are clearly evident.
Both $z_{c}$ and $y_{\rm c}$ will decay to zero (as required for
$\theta=\pi/2$), and their noise terms vanish at that point. By
contrast, the equation for $x_{\rm c}$ is independent of the others, and is
purely stochastic:
\beq
dx_{\rm c} = \sqrt{\gamma}dW(t)(1-x_{\rm c}^{2}).
\eeq

Clearly the equatorial pure states with $x_{\rm c}=\pm 1$ are
stationary solutions to this problem. Also, the system will tend to
one of these states. We can see this by calculating
\beq
d{\rm E}[x_{\rm c}^{2}] = \gamma dt{\rm E}\sq{(1-x_{\rm c}^{2})^{2}},
\eeq
which is always positive. That is, on average $x_{\rm c}^{2}$ always
increases. But it is also clear that
$x_{\rm c}$ has no preference to go to either of these states. Hence
they are not stable. The ensemble average $x$ is unchanging
under this evolution. Thus a perturbation which moves the state from
$x_{\rm c}=1$ to
$x_{\rm c} = 1-\epsilon$ say, will result in a proportion
$\epsilon/2$ of the states ending up at $x_{\rm c}=-1$, and a
proportion $1-\epsilon/2$ ending up at $x_{\rm c}=1$.

We have illustrated these features by showing two typical
trajectories in Fig.~\ref{fig:traj}(b). Once again, the initial
evolution is highly erratic, but the system reaches a fixed point
on a time scale of a few $\gamma^{-1}$. However, with the same initial
condition (the ground state), one trajectory ends up at $x_{\rm c}=1$
and the other at $x_{\rm c}=-1$.

\subsection{Non-unit-efficiency}

In the unit-efficiency case the stationary solution of the master
equation is (except for $|\theta|=\pi/2$), a pure state. This is
very special in that in means that every unraveling of the master
equation as a SSE or SME must end up in this same pure state also.
For non-unit-efficiency we have found the most pure stable state for
each $\theta$. In this case we must use a SME to unravel the master
equation, since the conditioned state will not be pure in general,
because of the inefficient detection.
Since the deterministic steady state is not pure (except for $|\theta|=\pi$),
 the quantum trajectories need not end up in this
state. Instead, the quantum state in an individual trajectory may
continue to evolve stochastically even when the system is in steady
state, and the equivalence to the deterministic evolution may hold
only on average. On the other hand, it is also possible that the quantum
trajectories do all end up in the deterministic steady state, since
we expect the conditioned state to be mixed anyway.

It turns out that with the optimal values of $\alpha$ and $\lambda$
defined in Sec.~IV, the actual behaviour is the first option
described above. That is, the system state continues to vary
stochastically in the long time limit, but is constrained so that the
time-averaged state is equal to the solution of the deterministic
master equation. We show this in Bloch space
Fig.~\ref{fig:bs3} for $\eta=0.8$ and $\theta=\pi/6$. We see that
 the amount of randomness in the system state in the long-time limit
 is quite large even for fairly high $\eta$.

This result suggests another question: would a different choice for
$\lambda$ be able to  reduce, or even eliminate, the
 randomness in the steady-state quantum trajectory, even though it
would necessarily be at the expense of the purity of the
deterministic stationary solution. [Recall that for a given
$\lambda$, $\alpha$ is still necessarily fixed by \erf{alpha}.]
To test this idea we tried choosing $\lambda$ based not
on  maximizing $r^{2}$ as in \erf{minr}, but on minimizing
\beq
N_{\theta}(\lambda) = \left|\left(\begin{array}{c}
-\sqrt{\gamma\eta}x_{\rm ss}^2+(\sqrt{\gamma\eta}+
{2\lambda}/{\sqrt{\eta}})z_{\rm ss}+\sqrt{\gamma\eta}\\
-\sqrt{\gamma\eta}x_{\rm ss}y_{\rm ss}\\
-(\sqrt{\gamma\eta}+{2\lambda}/{\sqrt{\eta}})x_{\rm
ss}-\sqrt{\gamma\eta}x_{\rm ss}z_{\rm ss}
\end{array}\right)\right|^{2}.
\eeq
That is, we minimize the noise terms in the SBE \erf{SBE1}. Note
that we have replaced the conditioned Bloch variables $x_{\rm c}$
{\em etc.} with the deterministic stationary solutions $x_{\rm ss}$
{\em etc.}, and that the dependence of these stationary
solutions on $\alpha$ and $\lambda$ add a further, implicit, dependence on
$\lambda$ to $N_{\theta}(\lambda)$. This is a sensible procedure if
the aim is realizable, and the noise in the solutions is reduced or
eliminated  so
that the conditioned states are approximately or exactly equal to the
deterministic stationary solution.

It turns out that this procedure cannot significantly reduce the
amount of steady-state randomness in the quantum trajectories below
that resulting from minimizing the deterministic stationary purity. In
fact, for all values of $\eta$ we considered, the variation of $\lambda$
(as a function of $\theta$)
based on minimizing the noise was indistinguishable by eye from that
based on maximizing the purity. This is not too surprising, but could
not have been predicted {\em a priori}.

\section{Conclusion}

We have given a rigorous analysis of the anti-decoherence feedback scheme
proposed by Hofmann, Mahler and Hess \cite{HofMahHes98}.
They proposed modulating the driving of a two-level atom
using the instantaneous homodyne photocurrent, in order to stabilize the atom
 in an arbitrary known pure state. We have shown that,
for detection efficiency $\eta=1$, the pure states thus
produced are stable. This is contrary to the conclusion of HMH, that
only pure states in the lower half of the Bloch sphere would be stable.
The one exception we found is for pure states on the equator. Although they
are
fixed points of the dynamics, they are not stable. A small
perturbation away from one fixed pure state
leads to a proportionally small fraction of the ensemble ending up in
the diametrically opposite pure state.

It is nevertheless possible to obtain an asymmetry between the upper
and lower halves of the Bloch sphere, reminiscent of the
conclusion of HMH, if one considers detection efficiencies less than one.
In this case, it is no longer possible to stabilize the system in a
given pure state, so
we choose the feedback and driving so that the solution of the master
equation (including feedback) is as close as possible to a given pure
state. We find that the purity (which measures this closeness) of
states thus produced decays to zero as $\eta$ decreases to zero,
for states in the upper half of the Bloch sphere. By contrast, those
in the lower half do not decay to zero. This is readily understandable
since in the absence of feedback (which is the situation which must prevail
when the detection efficiency goes to zero), the master equation with
driving alone has stationary solutions in the lower-half plane with
non-zero purity. The purity decays most rapidly with $\eta$
for states near the
equator, which is unsurprising given the instability of states on the
equator even for $\eta=1$.

In the non-unit-efficiency case, the state of the system conditioned
on the homodyne measurement results continues to evolve stochastically
even in the long-time limit, where the ensemble average evolution has
reached the desired most-pure state. Moreover, it seems that any
other choice of driving and feedback will result in more, not less,
randomness in the steady-state quantum trajectory.

Our results are significant in a number of ways. First, they show the
power of the quantum trajectory and master equation techniques developed in
Refs.~\cite{WisMil93b,WisMil94a,Wis94a}. Those techniques were
particularly useful for illuminating  subtle
questions regarding the stability of pure states, and for treating
inefficient detection.
Second, the physical system (the two-level atom) may one day find
application as a quantum bit in quantum information technology
\cite{WilCle98}. In that eventuality, the ability to stabilize the
atom against dissipation in an arbitrary (known)
pure state may be useful. Third, the system is a
simple but non-trivial example of quantum feedback in a nonlinear
system (the two-level atom). Thus the effectiveness of  feedback,
and in particular the influence  of non-unit-
efficiency detection on this effectiveness,
is of interest for what it may tell us about other
more complicated nonlinear systems.

In this last context, it would be of interest to also consider the
effect of non-Markovian feedback; that is, feedback with a time delay
or non-flat loop response function. This is much more difficult to
treat than Markovian feedback because the Lindblad master equations
derived in Refs.~\cite{WisMil93b,WisMil94a,Wis94a} do not apply.
Analytical solutions for non-Markovian feedback are possible for
linear systems \cite{WisMil94a,GioTomVit98}. For a nonlinear system
like the two-level atom, numerical simulations, or novel analytical
approaches, are necessary. This is an issue we plan to explore in
future work.

\section*{Acknowledgments}

This work has been supported by the Australian Research Council, the
University of Queensland, and the Department of Employment, Education and
Training, Australia. We would like to
 acknowledge useful discussions with Gerard Milburn.

\end{multicols}

\begin{references}

\bibitem{HauYam86}
H. A. Haus and Y. Yamamoto,
Phys. Rev. A {\bf 34}, 270 (1986).

\bibitem{YamImoMac86}
Y. Yamamoto, N. Imoto and S. Machida,
Phys. Rev. A {\bf 33}, 3243 (1986).

\bibitem{Sha87}
J. M. Shapiro {\em et al},
J. Opt. Soc. Am. B {\bf 4}, 1604 (1987).

\bibitem{WisMil93b}
H. M. Wiseman and G. J. Milburn,
Phys. Rev. Lett. {\bf 70}, 548 (1993).

\bibitem{Wis94a} 
H. M. Wiseman,
Phys. Rev. A {\bf 49}, 2133 (1994).

\bibitem{Pli94}
L. Plimak,
Phys. Rev. A {\bf 50}, 2120 (1994).

\bibitem{DohJac99}
A.C. Doherty and K. Jacobs,
Phys. Rev. A {\bf 60}, 2700 (1999).

\bibitem{WisMil94a} 
H.M. Wiseman and G.J. Milburn,
Phys. Rev. A {\bf 49}, 1350 (1994).

\bibitem{LieMil95}
A. Liebman and G.J. Milburn,
Phys. Rev. A {\bf 51}, 736 (1995).

\bibitem{MabZol96}
H. Mabuchi and P. Zoller,
Phys. Rev. Lett. {\bf 76}, 3108 (1996).

\bibitem{TomVit94}
P.~Tombesi and D.~Vitali,
Phys. Rev. A {\bf 50}, 4253 (1994).

\bibitem{HorKil97}
D.~B.~Horoshko and S.~Ya.~Kilin,
Phys. Rev. Lett. {\bf 97}, 840 (1997).

\bibitem{VitTomMil97} 
D.~Vitali, P.~Tombesi, and G.J.~Milburn,
Phys. Rev. Lett. {\bf 79}, 2442 (1997).

\bibitem{HofHesMah98}
H.~F.~Hofmann, O.~Hess, and G.~Mahler,
Optics Express {\bf 2}, 339 (1998).

\bibitem{HofMahHes98}
H.~F.~Hofmann, G.~Mahler, and O.~Hess,
Phys. Rev. A, {\bf 57}, 4877 1998.

\bibitem{Lin76}
G. Lindblad,
Commun. Math. Phys. {\bf 48}, 199 (1976).

\bibitem{Car93b}
H.J. Carmichael,
{\em An Open Systems Approach to Quantum Optics}
(Springer-Verlag, Berlin, 1993).

\bibitem{WisMil93a} 
H.M. Wiseman and G.J. Milburn,
Phys. Rev. A {\bf 47}, 642 (1993).

\bibitem{Gar85}
C.W. Gardiner,
{\em Handbook of Stochastic Methods}
(Spring\-er, Berlin, 1985).

\bibitem{WilCle98}
C.P. Williams and S.H. Clearwater,
{\em Explorations in Quantum Computing}
(Springer, New York, 1998).

\bibitem{GioTomVit98}
V. Giovannetti, P.Tombesi, and D.Vitali,
Phys. Rev. A, {\bf 60}, 1549 (1999).
\end{references}
\end{document}